\documentclass[12pt,preprint]{aastex}

\def\thn{{\thinspace}}
\def\sc{\scriptstyle}
\def\Msun{\hbox{$\thn M_{\odot}$}}
\def\Rsun{\hbox{$\thn R_{\odot}$}}
\def\Lsun{\hbox{$\thn L_{\odot}$}}
\def\etal{{{\it et al}.\ }}
\def\={\thn\thn=\thn\thn}
\def\halfline{{\null \vskip 7 truept}}
\def\tgs{{\thn \rlap{\raise 0.5ex\hbox{$\sc  {>}$}}{\lower 0.3ex\hbox{$\sc  {\sim}$}} \thn }}
\def\tls{{\thn \rlap{\raise 0.5ex\hbox{$\sc  {<}$}}{\lower 0.3ex\hbox{$\sc  {\sim}$}} \thn }}
\def\tll{{\raise 0.3ex\hbox{$\sc  {\thn \ll \thn }$}}}
\def\tgg{{\raise 0.3ex\hbox{$\sc  {\thn \gg \thn }$}}}
\def\tle{{\raise 0.3ex\hbox{$\sc  {\thn \le \thn }$}}}
\def\tge{{\raise 0.3ex\hbox{$\sc  {\thn \ge \thn }$}}}
\def\tl{{\raise 0.3ex\hbox{$\sc  {\thn < \thn }$}}}
\def\tg{{\raise 0.3ex\hbox{$\sc  {\thn > \thn }$}}}
\def\ts{{\raise 0.3ex\hbox{$\sc  {\thn \sim \thn }$}}}

\def\tp{{\raise 0.3ex\hbox{\small +}}}

\def\etal{\it et al.\ \rm}
\def\x{\ \ \ }
\def\by{\times}
\def\z{\ \ }

\input psfig.sty

\shorttitle{The Helium Flash}
\shortauthors{Dearborn, Lattanzio and Eggleton}

\received{2005 June 28}
\begin{document}

\title{3D Numerical Experimentation on the Core Helium Flash\\ of low-mass Red Giants}

\author{David S. P. Dearborn\altaffilmark{1}, John C. Lattanzio\altaffilmark{2} and Peter P. Eggleton\altaffilmark{1}}
\affil{$^1$Lawrence Livermore National Laboratory, 7000 East Ave, Livermore, CA94551, USA}
\affil{$^2$Monash University, Mathematics Department, Clayton, Victoria, 3168, Australia}
\email{dearborn2@llnl.gov, john.lattanzio@sci.monash.edu.au, ppe@igpp.ucllnl.org}

\begin{abstract}

We model the core helium flash in a low-mass red giant using 
Djehuty, a fully three-dimensional (3D) code. The 3D structures were 
generated from converged models obtained during the 1D evolutionary 
calculation of a 1$\Msun$ star.  Independently of which starting point we 
adopted, we found that after some transient relaxation the 3D model 
settled down with a briskly convecting He-burning shell that was not 
very different from what the 1D model predicted.

\end{abstract}


\section{Introduction}

The core helium flash is an important event in the life of most stars 
with a zero-age mass between about 1 and 2 $\Msun$; the minimum masses are 
a little lower for metal-poor stars. Since the work of Mestel (1952) 
and Schwarschild \& H{\"a}rm (1962) it has been clear that such stars 
ignite helium in a thermonuclear runaway situation, the helium flash, 
because the helium core is electron-degenerate at the time of 
ignition. Empirically, it is clear that this runaway is (usually) not 
a catastrophic affair, like a supernova explosion, because a whole 
class of stars, the horizontal-branch stars of globular clusters, is 
well explained by the survival of helium-flash stars in a long-lived 
state of core helium burning (Faulkner 1966). Nevertheless, attempts 
to compute the evolution during the flash have a confusing history: 
some calculations (both 1D and 2D) have predicted a rather severe 
explosion, and others (both 1D and 2D) a relatively benign though 
rapid ignition.

Most calculations until fairly recently have been 1-dimensional (1D) 
simulations, in which turbulent convection has been treated by a 
spherical averaging process based largely on the mixing-length 
concept. Deupree (1996), who gives a nice summary of earlier work, 
performed some 2D (axially symmetric) simulations. He found that 
the 2D estimates were critically dependent on approximations
made regarding eddy viscosity. We expect that by working in 3D we will 
not need to make such approximations, and we suggest that our results
bear out this expectation.

The Djehuty project of the Lawrence Livermore National Laboratory is 
an effort to model stars in 3D. Our ultimate aim is to be able to 
model an entire star, up to and including the photosphere; and indeed 
to generalize this to binary stars, including gas flows between them 
in, for example, a Roche-lobe-overflow situation. We are approaching 
this goal, but it is fairly easy to see that a star like the Sun, for 
instance, would require at least $10^{12}$ nodes inside it if there is to 
be adequate resolution near the surface. As computer power continues 
to increase this will no doubt become possible, but for the present 
we limit ourselves to about $10^8$ nodes. We therefore content ourselves 
with a simulation of the He flash that includes only the He core and 
the radiative portion of the envelope; we ignore the deep surface 
convection zone.

Apart from the intrinsic interest of the He flash, our other reason 
for pursuing this particular problem is that it potentially is a very 
good test for the stability and accuracy of a hydrodynamic code. This 
is because we have, as mentioned above, a rather good reason to 
anticipate what the outcome should be. We do not expect it to become 
a violent supernova-like event. In following an explosion, it is not 
easy to look at the outcome and say `that is clearly what should have been 
expected', even if it is what we expected. But in a non-explosive 
situation it is not difficult to compare, for instance, the heat flux 
actually carried by turbulent convection across a spherical shell 
with the expectation from a simple mixing-length model.

  In this paper we consider only non-rotating and non-magnetic cores,
but we believe that both these processes could be important and we 
hope to address them in a later paper.
In Section 2 we briefly outline the code; in Section 3 we describe (a)
our 1D input models and (b) our 3D output models. In Section 4 we
describe some issues and numerical tests regarding the stability of 
the calculation.

\section{Code Description}
Djehuty is a code designed to model entire stars in 3D. Developed 
at the Lawrence Livermore National Laboratory (LLNL), 
it operates in a massively parallel environment, and includes the 
basic physics necessary for modeling whole stars. Earlier 
descriptions of the Djehuty code can be found in Baz{\'a}n et al. 
(2003) and Dearborn et al. (2005), but a brief description will be 
provided here.

For a star, the Djehuty mesh is formed from 7 logically connected 
blocks of hexahedral cells of variable shape. There is a central cube 
of  $N\by N\by N$ cells, surrounded by six logical cuboids 
($N\by N\by L$). One of 
the two $N\by N$ faces of each cuboid is attached point by point to the 
face of the central cube.  The other $N\by N$ face of each cuboid is 
mapped to lie on a spherical surface forming the outer boundary. The 
cuboids are then `morphed' into wedge shapes with surfaces on the 
long (L) radial axis transitioning from planar to spherical.  The $N\by L$ 
faces are similarly attached to adjacent cuboids in a point-by-point 
fashion. An exploded version, with the logical structure of the 
central cube and surrounding cuboids, is shown in Fig. 1 with 
$ N = 50$ and $L=100$; the lower part of Fig. 1 shows two of
the blocks connected together and in physical space, with $N=35$
and $L=70$.
\begin{figure*}
\vskip -0.6truein
\centerline{\psfig{figure=f1.eps,height=7.5in,bbllx=0pt,bblly=0pt,bburx=570pt,bbury=760pt,clip=}}
\caption{(Upper) An exploded view of the mesh structure, in logical 
space. (Lower) Blocks 0 and 4 in coordinate space (here $N=35$ and $L=70$).
Block 0 is a great deal smaller than Block 4.}
\end{figure*}

This mesh structure allows reasonable azimuthal resolution without 
the core convergence problem (tiny zones and tiny timesteps at the 
center) endemic in spherical coordinate systems.  In the core itself, 
the cells are nearly rectangular, and as the radius grows successive 
surfaces become more spherical, matching the potential surfaces as 
well as properties like temperature and pressure.   The mesh can 
encompass an entire star with free outer boundaries, as was done by 
Dearborn et al. (2005), or a portion of the star with various fixed 
boundary conditions, as will be done here. For reasons discussed 
later, in these calculations we will locate the outer boundary at the 
inner edge of the red giant's deep convective envelope. 

The initial 3D 
structures are generated from models produced by a 1D hydrostatic 
stellar evolution code.  This 1D code was used as a platform to 
test the physics (equation of state, nucleosynthesis, etc.) 
incorporated into Djehuty, as well as to provide structure 
information for constructing 3D models.  Djehuty reads the 1D stellar 
models, mapping the defining physical parameters on to a 3D spherical 
grid at any given stage of evolution. The radial structure of the 1D 
mesh is used as a guide in scaling the 3D mesh, so that regions with 
steep gradients, like thin burning shells, are resolved.  Scalar 
quantities like pressure, temperature and composition are assigned to 
the cell centers, and vector quantities (position, and velocity) are 
assigned to nodes. The position of the cell center is the direct 
average of the 8 radius values of the nodes surrounding it.

The 1D code carries 6 elements ($^1$H, $^3$He, $^4$He, $^{12}$C, 
$^{14}$N, and $^{16}$O) 
directly, with the remainder assumed to be $^{24}$Mg.  This set of elements 
was selected to allow accurate tracking of the principal energy-generation 
reactions for hydrogen and helium burning over a broad 
range of masses and evolutionary stages.  The 7-element set includes 
the triple-alpha reaction, as well as the alpha captures on carbon and 
nitrogen, to provide the energy production, but the 1D code does not 
follow $^{18}$O so that the capture on nitrogen is assumed to be a 
2.5$\alpha$ capture to the heavy element remainder ($^{24}$Mg).

The 3D code is capable of operating with the same element set as the 
1D code, but for this calculation the 3D code followed a 21-element 
set for a more accurate definition of the helium burning results. 
The elements followed are
$^{1}$H, $^{3}$He, $^{4}$He, $^{12}$C, $^{13}$C, $^{13}$N, $^{14}$N,
$^{15}$N, $^{15}$O, $^{16}$O, $^{17}$O, $^{18}$O, $^{17}$F, $^{18}$F, 
$^{19}$F , $^{20}$Ne, $^{22}$Ne, $^{24}$Mg, $^{28}$Si, $^{32}$S and 
$^{56}$Ni.
In the 7-element mode, the rate equations are identical in the 1D and 
3D codes. When the 21-element set is used, the 3D code has those 
rates as well as additional rates and couplings.  The 21-element set 
is connected with an extensive set of nuclear reaction rates 
including hydrogen, helium, carbon and oxygen burning reactions, as 
well as a small NSE (nuclear statistical equilibrium) approximation 
(Timmes et al. 2000). Tests using 
the two networks for a range of fixed conditions representative of 
static hydrogen and helium burning have shown that the networks are 
well matched in their energy production rates. Coulomb screening 
(Graboske \etal 1973) and neutrino cooling (Itoh 
\etal 1989, 1992, with errata) are implemented in both codes.

The 1D code uses the Eggleton (1971) approach for implicit, adaptive
mesh adjustment, with simultaneous implicit solution for the mesh {\it
and} the composition along with the structure, permitting the code to
continue smoothly and efficiently
through radical structural adjustments.  The differencing is done in
such a way as to permit a solution by the technique pioneered in the
stellar-structure context, and for the structure variables only,
by Henyey et al. (1959), in which some boundary
conditions are central and some are at the surface.

In 1D, convection is always modeled by an approximate process.  In
our 1D code, convective energy transport is treated with a standard
mixing-length approach, and element mixing is modeled as a diffusion
process using a second-order differential equation with advection
and nuclear-burning terms for each element.  Limits of convective
regions are determined with a Schwarzschild stability criterion. 

The 3D code has no such approximations.  Material moves when there is a
force.  We have striven for sufficient spatial resolution to ensure
that the larger convective eddies that arise spontaneously are
reasonably resolved. It is these larger eddies that tend to carry
most of the heat. The code provides options for `sub-grid modeling',
i.e. estimates of how much heat might be carried by unresolved
small-scale eddies, but we did not use them. Our mesh allowed
10 -- 20 cells (in each dimension) within the major eddies, and
we feel that this is sufficient for their resolution.
This feeling is supported by comparing runs with a mesh that had 
twice as many cells.  The energy transport, as indicated by the 
helium burning rate, was effectively the same.  While the resolution 
used here appears sufficient for modeling the physical behavior of 
interest (stability of the helium burning shell), we do not claim to 
have captured the smaller scale eddies through which the kinetic 
energy of the turbulent flow actually merges into the thermal field.

Beyond the modest resolution, the small-scale end of the turbulence 
spectrum is terminated by the ALE (Arbitrary Lagrange-Eulerian) 
hydrodynamic method, used to allow a Lagrangian representation to 
survive in a sheared region. When motion causes the
Lagrangian mesh to be sufficiently distorted, an Eulerian re-map step,
or more precisely an interpolation step performed in a manner similar
to what is used in a numerical approximation to advection,
can mix the composition of adjacent cells while smoothing the mesh
structure.
In stable regions, or regions of large-scale coherent motion where
the Lagrangian mesh is modestly deformed, no remapping is necessary,
and the code is essentially Lagrangian.  In regions where shear
develops, it is necessary to relax the mesh, and permit material to move
between cells. The transition to an Eulerian result is smooth and
accurate.

Such a code incorporates in effect two kinds of articial viscosity. Any
finite-difference scheme, Lagrangian or Eulerian, implies artificial
viscosity through approximating derivatives as finite differences. In
addition, the remap step which is applied from time to time to prevent
major distortion of the (normally) Lagrangian nodes will also introduce
a form of numerical diffusion or viscosity. We rely on these, and these
alone, to prevent the creation of small-scale (unresolvable) eddies
which would normally be driven by the spectrum of eddies that {\it are}
resolved, and which are liable to destroy the calculation if unchecked.

The ALE scheme implemented here was tested by Pember \& Anderson 
(2001), who ran a number of standard hydrodynamics tests comparing it 
to an Eulerian high-order Godunov scheme.  They found the accuracy of 
the two schemes to be generally equivalent, and that the ALE scheme was much 
improved over an earlier comparison reported by Woodward \& Colella 
(1984).  A predictor-corrector formalism promotes hydrodynamic 
accuracy, and time centering for other physical processes.  The 
hydrodynamics step is second-order accurate in both time and space.

Both the 1D and 3D codes use the same analytic equation of state 
developed by Eggleton et al. (1973), as updated by Pols et al. (1995), 
which provides continuous thermodynamic 
derivatives for hydrodynamic consistency.  It includes molecular 
hydrogen binding, as well as ionization of the light elements, and 
can reproduce tabulated values of Rogers \& Iglesias (1992) to much better 
than 1\% accuracy for the entire range of conditions expected in stars 
between 0.7 to 50.0 $\Msun$, over their whole evolution.

The code operates with separate matter and radiation temperatures 
integrated in flux-limited diffusion equations.  These equations 
include energy source and sink terms that link the radiation 
equation to the hydrodynamic energy and momentum equations, as well 
as to nuclear energy production and neutrino losses.   Planck and 
Rosseland mean opacities are derived from the Opal library at LLNL, 
and Alexander opacities for the lower temperatures (Alexander \& 
Ferguson 1994, Rogers \& Iglesias 1992). Conduction coefficients for 
electron heat flow are from the conductive opacities tables of
Hubbard \& Lampe (1969), as modified by Itoh \etal (1983). 

The hydrodynamics and the gravitational potential are purely Newtonian; 
the latter uses integration of a mass-radius relation to generate a spherical 
potential which approximates the gravity. While this is adequately accurate 
for the problem considered here, we are developing a 
multipole approximation to the gravitational potential for use in 
less spherical systems.

As a final note on the 1D modeling of a helium flash, the conditions 
make numerical stability a challenging problem.  To this end, the 1D
code was made fully (64-bit) double-precision, the derivatives necessary for 
this solution were all done analytically, and the convergence criteria 
for the Newton-Raphson solver were set tightly (usually 1 part in $10^7$,
although 1 part in $10^6$ seemed adequate in relatively easy phases).

\section{The Helium Flash}
\subsection{One Dimension}
Each of the 3D simulations done here started from a basic model 
generated as part of the evolutionary sequence of a $1\Msun$ Pop. I star. 
It assumed an initial hydrogen mass fraction of 0.7, and metals abundance 
($Z$) of 
0.02.  The model had 750 zones, and a mixing length ratio of 1.8. With 
these initial conditions, it did not match our best solar model, 
but was close ($L=1\Lsun$ at $4.58\thn$Gyr, and R=1.02$\Rsun$).

Modeling began with a fully convective pre-main sequence structure 
having no nuclear reactions. The 1D evolution passed through the main 
sequence and the giant branch, to the helium core flash, and finally 
to core helium burning (Fig. 2).  The helium core flash occurs at the tip 
of the giant branch when the model is $12.2\thn$Gyr old, and has a radius 
of 175$\Rsun$. At this point the core mass is 0.472$\Msun$ with a radius of 
0.026$\Rsun$.  Before helium ignition all of the 2670$\Lsun$ is derived 
from hydrogen burning in a thin shell.  Surrounding the hydrogen 
burning shell is a radiative region that contains only about 0.002$\Msun$, 
but whose radial extent exceeds $\ts\Rsun$ or 40 core 
radii.  Over this radiative region, the density drops by nearly 5 
orders of magnitude.
\begin{figure*}
\vskip -0.5truein
\centerline{\psfig{figure=f2.eps,height=7.2in,bbllx=0pt,bblly=0pt,bburx=580pt,bbury=750pt,clip=}}
\caption{The 1D evolution of our 1$\Msun$ model in the theoretical HRD.  
Pre-flash evolution is in blue, and post-flash evolution is in red. 
During the later stage of evolution to the horizontal branch there
were two major oscillations (`mini-flashes'), which appear as a somewhat
broadened part of the red track.}
\end{figure*}

\par During the entire 1D evolution mass loss by stellar wind was {\it not}
included. It is well-known that some mass loss, probably before the
helium flash as well as after it, is to be expected. However it is also
well-known that the helium {\it core} on the approach to the helium flash is 
rather 
little affected, provided only that the mass loss does not strip the star right
down to the core. Since the core at the helium flash was $0.472\Msun$, the
star would have had to lose somewhat more than half its mass. This is not
out of the question, but it is probably an estimate on the high side. If
either (a) the star lost only $0.45\Msun$, or alternatively (b) started
at say $1.3\Msun$ and lost $0.75\Msun$, the core at the flash would be
very much the same as the one we obtained. Thus we feel that the `default
option' of no mass loss is a very reasonable simplification.

In the helium core, neutrino cooling reduces the central temperature, 
shifting the initial helium burning region off-center.  Energy production 
from helium burning starts in a narrow region at about $0.18\Msun$, or 
$0.008\Rsun$, from the center.  As the energy production rate from 
helium burning becomes significant, a convective shell is developed 
to transport the energy.  The 1D hydrostatic code uses a mixing-length 
approach that has no time dependence for the start-up of 
convective heat transport (although because mixing of composition is
modeled by a diffusion process the response of composition is {\it not}
instantaneous).  As is the common practice, the 1D code uses a stability 
criterion 
to determine where convection is necessary.  As we will show in our 
3D modeling, the timescale for developing convection, and its 
efficiency at removing energy from the very thin burning region, is 
critical to the stability of the simulations, as well as to the 
timescale for the flash.
\def\sc{}
\def\y{\ \ \ \ \ \ \ \ \ \ \ \ \ \ \ \ \ \ \ }
\def\x{\ \ \ \ \ }
\def\Lscun{\hbox{$\sc  \thn L_{\odot}$}}
\halfline
{\centerline{Table 1. The time difference $\sc t-t_{\rm peak}$ (days)}
\vskip -0.1truein
\baselineskip=14truept
$$\vbox{\halign{\hfil#&#\hfil&#\hfil\cr
$\sc \log(L_{\rm He}/\Lscun)\x$&pre-peak&p
ost-peak\cr
\z         &                   &                         \cr
\z	9\y&	-3.28          &        4.38             \cr
\z	8\y&	-14.23         &        28.47            \cr
\z	7\y&	-67.16         &        202.57           \cr
\z	6\y&	-375.95	       &        1.5$\sc \by 10^3$\cr
\z	5\y&-2.2$\sc \by 10^3$\y&	1.2$\sc \by 10^4$\cr
\z	4\y&-1.4$\sc \by 10^4$\y&	9.8$\sc \by 10^4$\cr
\z	3\y&-7.0$\sc \by 10^4$\y&	5.2$\sc \by 10^5$\cr
\y	2\y&-4.9$\sc \by 10^5$\y&	1.3$\sc \by 10^6$\cr
}}$$}
While the ignition of helium is fast by the standards of stellar 
evolution, it is slow compared to hydrodynamical time scales (the 
sound travel time across the core, a few seconds).  From the time that 
the energy 
production rate from helium burning reaches $100\Lsun$, it takes about 
1000 years to reach $1000\Lsun$, and another 140 years to reach $10^4\Lsun$. 
The energy production rate peaks above 3$\by 10^9\Lsun$, for a period just 
over 30 hours, and the model spends about 7 days producing energy 
above $10^9\Lsun$ (Table 1).
During the height of the energy production, nearly all of this energy 
is used to expand the outer portion of the core (Fig. 3), changing its 
thermal structure.  As the core expands, the energy produced by 
hydrogen burning at the edge of the core falls.  By the time of peak 
energy production, 
hydrogen burning produces only about $200\Lsun$.  It is only $0.03\Lsun$ a 
year after the peak of the helium flash.  About 8000 years after the peak 
of the helium core flash, the hydrogen burning shell has begun to 
re-establish itself at a lower level, and the energy production rate 
by both hydrogen and helium burning are near $20\Lsun$.

\begin{figure*}
\vskip -0.5truein
\centerline{\psfig{figure=f3.eps,height=7.2in,bbllx=0pt,bblly=0pt,bburx=550pt,bbury=750pt,clip=}}
\caption{In the 1D model, the time history of the energy production rate 
(in $\Lsun$) from helium burning is shown in red.  The absolute value of 
the rate of change in the thermal energy of the model is shown in blue.}
\end{figure*}

Because the energy produced by the hydrogen burning shell drops, it is 
unable to support the very large radius.  As the energy stored in the 
envelope leaks out, the radius decreases.  In the 15,000 years 
following the peak of the helium flash, the radius drops to about
$20\Rsun$, near to the value that will be supported by core helium burning.

The initial burst of helium burning energy is absorbed in a 
convective shell, and over-expands the core.  This quenches the 
helium burning (Fig. 4), and the convective shell stabilizes. 
At this point, the core structure includes a central region of about 
$0.16\Msun$ ($0.008\Rsun$) where no helium burning has occurred 
(yet), surrounded by a thoroughly mixed shell region.

\begin{figure*}
\vskip 1.5truein
\centerline{\psfig{figure=f4.eps,height=3.8in,bbllx=0pt,bblly=0pt,bburx=570pt,bbury=400pt,clip=}}
\caption{In the 1D model, following the initial helium flash there are two 
subsequent mini-flashes as the core relaxes to a fully convective 
structure.}
\end{figure*}

Eventually the model radiates the excess thermal energy that it has 
absorbed, and attempts to re-establish helium burning. Following this 
initial flash, the mass fraction of $^{12}$C averaged over the entire core 
is 0.025.  On a timescale of about $2\by10^5$ years, the model experiences 
two more mini-helium-flashes.  These mini-flashes appear in the HR 
diagram (Fig. 2) as a couple of narrow loops loops near the ultimate core 
helium burning luminosity.  Following the second mini-flash, the 
fluctuations are small enough that the associated convective region 
is never extinguished.   Ultimately, about $2\by10^6$ years after the 
peak, the center becomes completely mixed, and the average mass 
fraction of $^{12}$C is 0.035.

\subsection{Three Dimensions}
The entire helium flash duration is much too long to be followed by a 
Courant-limited hydrodynamics code, but such a code is essential to 
study the timescale on which convection develops to stabilize the 
nuclear burning.  This development is the fundamental determinant for 
the duration of the flash, and the type of star that emerges from it. 
3D modeling permits a direct simulation of the 
convection process, as well as numerical testing of the sensitivity 
to perturbations or limitations (such as resolution).

For this study, 1D models were stored at intervals along the 
evolutionary track, and three of them were selected to study in four 
3D simulations (Table 2).  As noted in the code description, Djehuty 
is capable of modeling entire stars, but for this problem, it is the 
helium core and hydrogen shell region that are of interest.  It 
contains nearly half of the mass of the star, but only $10^{-12}$ the 
volume.  The vast bulk of the star is in the convective envelope.  To 
concentrate our effort on the portion of the star that was pertinent 
to this study, we modeled only a portion of the star, selecting a 
radius that is in the radiative region below the convective envelope. 
This outer radius was held fixed in temperature and size (Fig. 5). 
Even here we were modeling a region that had nearly 30 times the 
radius (27,000 times the volume) of the helium core, and across which the 
density drops by nearly 5 orders of magnitude.
\begin{figure*}
\vskip -0.7truein
\centerline{\psfig{figure=f5.eps,height=7.5in,bbllx=0pt,bblly=40pt,bburx=570pt,bbury=650pt,clip=}}
\caption{For model E4, the mesh is shown on a plane slice through the star, slightly off
center. The color background is scaled to the temperature, blue cold and red hot.
The peak temperature is about $17\thn$Kev. The high temperature ring 
in the bottom left panel is the base of the helium burning shell.  The central 
mesh block, block 0, is contained inside this region.
In the lower two panels, some closed curves that form squarish figures with
rounded corners are artefacts of the visualisation code, when it is required to
visualise the mesh rather than a variable such as temperature. They arise because a
plane surface cuts through the sphere, and this surface does not, as a rule, cut exactly
through any of the meshpoints. Neighboring meshpoints and lines are projected on to
it, but where the nearest meshpoints happen to be equidistant from the surface on
opposite sides this artefact is produced. It is a kind of interference pattern.}
\end{figure*}

One of the three models, E4, was selected to be a relatively benign 
case with an energy production rate from helium burning near $10^4\Lsun$. 
This model represents the state of the star more 
than half a century prior to the peak energy production rate, and 
less than $10^{-5}$ of that rate. The other two models were chosen 
to be near (E8) or at (E9) the peak energy production rate. 
Information on the selected models is given in Table 2.
All the 1D starting models are available electronically, on request.

\vfill
\eject

\def\a{{\bf a}}
\def\v{{\bf v}}
\def\mv{\vert\v\vert}
\def\={\ =\ }
\def\pr{^{\prime}}
\def\dt{\delta t}
\def\dm{\delta M}
\def\vtr{V_{\rm treacle}}
{\centerline{Table 2. Parameters for the Four Simulations }
\vskip -0.1truein
\baselineskip=14truept
$$\vbox{\halign{\hfil#&\hfil#&\hfil#&\hfil#&#\hfil&#\hfil\cr
       &\x              &\x    Mesh           &\x              &\x Run      &\x$R_{\rm fixed}$\cr
Model  &\x$L_{\rm He}  $&\x	(mega-zones)&\x Processors   &\x Time(s)  &\x(cm)    \cr
E4     &\x        $10^4$&\x	0.39\x	    &\x24\x	     &\x4694      &\x6.0$\by10^{10}$\cr
E8     &\x	$2\by10^8$&\x	0.39\x      &\x24\x	     &\x6605      &\x6.5$\by10^{10}$\cr
E9     &\x	$3\by10^9$&\x	0.86\x	    &\x31\x	     &\x3665	  &\x8.2$\by10^{10}$\cr
E9$\pr$&\x	$3\by10^9$&\x	1.32\x	    &\x62\x	     &\x1917	  &\x8.2$\by10^{10}$\cr
}}$$}

In doing this suite of simulations we tried various resolutions 
(meshes with 0.39, 0.86, and 1.32 million zones) and tested different 
approaches to settling the transients associated with imperfections 
in the 1D-to-3D mapping process. Despite the efforts to match the 1D 
and 3D codes, the mapping process results in small deviations from 
hydrostatic equilibrium.  While the resulting motion should decay in 
the fullness of time, stellar interiors are remarkably good 
oscillators, and the motion can persist for quite a long time.  In 
effect, the lack of exact hydrodynamic equilibrium in the initial 
discretized model creates artificial pressure waves that can bounce 
around the model for many crossing times. The situation is made worse 
by the fact that as waves move towards the surface, down the 
considerable density gradient, they can reach the force of a tsunami.

Among the settling options is one (Zerovel) which is simply to set to zero 
the accumulated 
velocities, usually in outer regions where the wave has 
gathered strength.  A less intrusive option is to introduce, 
temporarily, a velocity limiter ($\vtr$).   This speed limiter is a 
very severe kind of artificial viscosity, as follows. Suppose that 
the acceleration computed at a particular node is $\a$. Then the 
equation of motion is taken to be
$${D\v\over Dt}\=\a \ \ {\rm if}\ \ \mv\tle V_0\pr\ ,\ \ \ {\rm but}\  \ \ 
{D\v\over Dt}\= \a-\v \thn{\mv-V_0\pr\over{\dt\mv}}\ \ 
{\rm if}\ \ \mv\tge V_0\pr\ .\eqno(1)$$ 
Here $\dt$ is the timestep and $V_0\pr$ is a critical (`treacle') speed.

 The speed limiter has been implemented in two forms.  In the form used 
here,  $V_0\pr$ is a user-specified constant, but it is also possible to 
select an option in which it scales with the node mass:
$$V_0\pr=V_0\thn\left({\dm\over\dm_0}\right)^{1/2}\ ,\eqno(2)$$
where $\dm$ is the mass of a cell, $\dm_0$ is the largest cell mass in the
star (usually near the center), and $V_0$ is a specified value. Either option 
effectively prevents the velocity from becoming much larger than $V_0\pr$ 
and in a smooth manner that does not lead to near-discontinuities. 

The constant $V_0\pr$ option was used here, because the drop in density though
large is not as large as in a whole star, where the variable option
is currently being tested. The constant option 
provides a well-defined user constraint that minimizes its impact. 
In the second option, the dependence on $\dm$ makes the damping more vigorous
near the surface where $\dm$ is small, and prevents the artificial pressure
waves from becoming large towards the outer layers. After some lapse of time, 
determined by trial and error, $V_0\pr$ is removed, i.e. is effectively raised 
to infinity, so that the correct equation of motion is solved from then on.

During the settling-down process we expect pressure waves to radiate
more-or-less spherically from the core. Given that we have an artificially
fixed outer boundary, we might expect them to bounce back, and cause further
trouble. However, the `treacle' viscosity which we described above seems
to be effective at damping them to insignificance even before they reach
the outer boundary.

For the helium flash problem, this settling option also provides an 
opportunity to perform numerical experiments on the sensitivity to 
convection.  Setting it to a value where only the very fastest nodes 
were affected allowed a quantitative assessment of the importance of 
convective efficiency. The result of such a numerical experiment is 
shown in Fig 6, plotting the time history of the energy 
production rates of models E4 and E8.
\begin{figure*}
\vskip 0.5truein
\centerline{\psfig{figure=f6a.eps,height=3.0in,bbllx=0pt,bblly=0pt,bburx=600pt,bbury=360pt,clip=}}
\centerline{\psfig{figure=f6b.eps,height=3.0in,bbllx=0pt,bblly=0pt,bburx=600pt,bbury=350pt,clip=}}
\caption{The upper panel (E4) shows the impact of using 
`Zerovel' as well as a constraint on the speed.  In the lower panel 
(E8), `Zerovel' was not used.  Here, a velocity limiter of 5 km/s 
allowed rapid settling from any imbalance from hydrostatic 
equilibrium.  In both cases, the speed limiter stabilized the energy 
production rate at a higher value than occurred without a limit.}
\end{figure*}

Both simulations show an initial spike in energy production 
associated with the lack of velocity information for the initial 
model.  Without convective motion, the temperature of material in 
the helium ignition region climbs swiftly. In model E4, the energy 
production rate reaches $1.8\by10^4\Lsun$ after only 0.3 seconds.  This 
layer rapidly expands, becoming Rayleigh-Taylor unstable, and 
initiates convection.  As buoyant fingers of material begin to rise, 
and are replaced by cooler material from above, the temperature is 
stabilized and the energy production rate returns to the expected value. 
Similarly, in model E8, there is an initial spike to $3\by10^8\Lsun$, and 
with the onset of convection the rate returns to the expected value 
after about 20 seconds.

For our first 3D simulation, E4, we experimented with the Zerovel 
approach to settling the initial structure.  In the first few hundred 
seconds we attempted a series of Zerovel tests in which the 
velocities outside the convective shell were set to zero.  Each of 
these perturbations resulted in a small energy spike, and the effort 
was abandoned.  Through this time, and out to approximately 800 
seconds, the speed limiter was set to 1 km/s.  Over this period, the 
energy production rate stabilized at about $6000\Lsun$.  When we removed 
the speed limit entirely, convection near the hottest spots could 
operate at a higher speed, and the energy production from helium 
burning dropped to about 3000$\Lsun$. Subsequently, it increased smoothly to 
4000$\Lsun$ after about an hour.  A stable but very thin convective 
shell had developed that was within the region predicted by the 1D 
code.  The energy production rate was slightly lower, but appeared to 
be approaching the original 1D model.

With the experience gathered from modeling E4, we moved to simulate a 
much more energetic structure, E8.  This model represents a star less 
than 15 days pre-peak, with an energy production rate from helium 
burning that is just over $10^8\Lsun$.  The velocity limiter was set to 
provide a speed limit near 5 km/s in the core, and we left it active 
for about 1400 seconds after startup.  At the imposed speed limit 
this was time for nearly 2 complete turnovers of the convective 
shell, though relatively little material was actually moving fast 
enough to activate the limiter.  The result of this speed limit was a 
convective shell that had a stable energy production rate of about 
$7\by10^7\Lsun$.  The speed limiter significantly impacted the energy 
production rate, because the high velocity nodes were systematically 
associated with the warm spots where the bulk of the energy 
production occurs.

After 1400 seconds, the speed limit was repealed, and in less than a 
turnover time, the luminosity dropped to about $3\by10^7\Lsun$. The model 
was then followed for an additional 4600 seconds, sufficient for many 
convective element turnovers. The final energy production rate here 
was about a factor of 3 below the 1D hydrostatic model value. This 
is the largest difference seen between the 1D and 3D models. Given 
the tremendous sensitivity of the energy production rate to the 
precise structure, we took the result to be acceptable, but it is 
certain that the long-term application of the velocity limiter 
has the effect of over-producing energy, expanding the structure. 
It is possible that 
the forced structure change led to the lower final luminosity.

\begin{figure*}
\vskip 0.5truein
\centerline{\psfig{figure=f7.eps,height=5.0in,bbllx=0pt,bblly=0pt,bburx=570pt,bbury=460pt,clip=}}
\caption{Comparing the composition structure of models E8 and E9, 
shows the hydrogen burning shell to be expanding at about 3.5 m/s. 
At the peak energy production rate, the convective shell extends from 
about $0.16\Msun$ to $0.43\Msun$ just below the hydrogen burning shell.}
\end{figure*}

The next model E9 considered in 3D represented the state of the star 
very near the peak of the energy production rate, near $3\by10^9\Lsun$. The 
1D model spends about 31 hours at this high-energy production rate 
before the rate begins to decrease. The composition structures for 
the $10^8\Lsun$ model and this one are shown in Fig. 7. Over the 
intervening days between these models, the whole core (as defined by 
the hydrogen burning shell) has expanded with an average speed of 2.5 
m/s, and the helium convective shell has developed out to $0.43\Msun$ 
(from $0.0085\Rsun$ to $0.0195\Rsun$).  As discussed in the 1D section, 
this expansion has caused the energy production from hydrogen burning 
to drop tremendously.

Because model E9 is very near the peak energy production rate, the 
start-up transient that occured while convection established itself 
was thought to be most likely to result in anomalous behavior (core 
disruption, core/envelope mixing, ...). Here the initial energy spike 
reached nearly $5\by10^9$\Lsun, and again quickly dropped to the expected 
value (Fig. 8).  The initial value of the speed limiter was set to 
$10\thn$km/s, 
and was removed after less than 100 seconds. We again attempted to 
apply the zerovel option (between 300 and 400 seconds), but by 400 
seconds simply left the simulation alone with no {\it ad hoc} options. 
This resulted in a structure that produced energy from helium burning 
near $1.8\by10^9\Lsun$ for the hour simulated.

This model was run twice, once with a 0.86 mega-zone mesh, E9, and 
again with 1.36 mega-zones, E9$\pr$.  The higher resolution (larger mesh) 
run was started the same speed limiter but it was removed after only 
50 seconds.  In E9$\pr$, no attempts were made to use the zerovel option. 
Except for the short period where the zerovel was tested in E9, the 
evolution of the energy production rate was in seen to be agreement 
in the separate simulations (Fig. 8).

\begin{figure*}
\centerline{\psfig{figure=f8.eps,height=5.0in,bbllx=0pt,bblly=0pt,bburx=570pt,bbury=460pt,clip=}}
\caption{ In both E9 (red) and E9$\pr$ (blue), there is an initial 
transient associated with the onset of convective motion, but this
quickly settles to a stable value near that of the 1D model.  The 
more refined mesh, E9$\pr$, produces a nearly identical result.}
\end{figure*}

Both simulations show small fluctuations in the instantaneous energy 
production rate as hot spots occur and are quenched by expansion and 
plume generation.  The fluctuations appear to be small in both 
simulations, but they are slightly smaller in the higher resolution 
case.  The behavior brings into question the overall stability of the 
lower resolution model against the hydrodynamic fluctuations that 
lead to hot spots.  To test this, a numerical experiment was performed 
on the lower resolution model, E9.  The temperature of 18 contiguous 
zones was artificially increased from their original values (near 16 
Kev, or $1.9\by 10^8\thn$K) to a value of 25 Kev.  The peak temperature 
variation seen in 
the unperturbed models was less than 100 ev, so a 9 Kev increase was 
a tremendous perturbation.  Nevertheless, the model demonstrated that 
it was stable against such fluctuations.  After the large temperature 
perturbation was artificially introduced, there was a spike in the 
total energy production rate which increased by a factor of 2 for about one 
second.  Following the spike, there was a trace of excess energy 
production for another 5 seconds as a rapidly rising plume was 
initiated (Fig. 9).

\begin{figure*}
\centerline{\psfig{figure=f9a.eps,height=4.0in,bbllx=0pt,bblly=0pt,bburx=570pt,bbury=460pt,clip=}}
\centerline{\psfig{figure=f9b.eps,height=4.0in,bbllx=0pt,bblly=0pt,bburx=570pt,bbury=460pt,clip=}}
\caption{Run E9 perturbed. Upper: the energy production rate from an artificially introduced 
hotspot shows local expansion and plume formation to be a robust 
stabilizing force. Lower: the plume forms a mushroom cloud as 
seen in an $^{18}$O contour (constant mass fraction).}
\end{figure*} 

  Within 20 seconds, that plume had risen farther and faster than any 
of the normal plumes, and the energy production rate had returned to 
the pre-hotspot level. This is strong evidence that even the 
coarser resolution used was stable against the normal temperature 
fluctuations in the convective shell. The $10^8\Lsun$ model (E8), was also 
tested for stability against a range of volume and temperature 
perturbations that far exceeded the natural fluctuations that were 
seen to occur naturally, and again proved stable.

As described above, the initial 3D models had no velocity 
information, and the initial motion occurs as a result of a strong 
burst of energy production in a thin layer that is at the bottom of 
the convective shell of the 1D model.  As this layer expands, the 
energy production rate decreases, but the layer becomes 
Rayleigh-Taylor unstable.  The result is a pattern of rising and 
falling areas that shows mesh imprinting associated with the small 
deviations from sphericity.  In the $10^9\Lsun$ model (E9) this patterning 
persists as plumes rise and settle for over 20 minutes of the hour 
modeled.  The pattern gradually dissipates, and well before the end 
of this run, the plumes appear to occur randomly with no special 
connection to the mesh structure.  

In Fig. 10, a contour is shown with 
a fixed mass fraction of $^{18}$O is.  As the 1D model did not track this 
isotope, all of it is produced in the 3D model, and it serves as a 
tracer of element production and convective distribution in the 
helium burning shell. Although some articial mixing is introduced
through the occasional Eulerian remapping of the Langrangian mesh,
this effect is small compared with the genuine changes of nuclear
evolution.

\begin{figure*}
\centerline{\psfig{figure=f10.eps,height=6.0in,bbllx=0pt,bblly=0pt,bburx=500pt,bbury=500pt,clip=}}
\caption{For run E9, the outer 1D convective limit is shown on a slice through 
the center of the star.  That plane also shows the local velocity 
vectors.  Superimposed on this slice is a 3D contour of fixed $^{18}$O 
mass fraction.}
\end{figure*}

Fig. 10 also shows the velocity vectors lying in a plane through 
the center of the star, and a circle showing the limit of convection 
in the 1D model.  In the outer areas, the velocity vectors show a 
persistent azimuthal ringing that has some mesh pattern in it.  This 
ringing does not result in mixing, and makes simple speed tracking 
undependable as an indicator of the operation of convection.  In the 
neighborhood of the $^{18}$O surface shown, the velocity vectors show a 
very different behavior. Here they show patterns of rising and 
falling regions as plumes develop, cool, mix, and fall.  Over the 
hour followed in model E9, substantial mixing occurred over the inner 
half of the expected region, and was slowly moving outward.

An alternative illustration of this mixing is shown in Fig. 11, a color 
plot showing the distribution of $^{18}$O on a slice through the star 
(run E9). The 
original convective shell is defined by $^{12}$C contours where this 
isotope has a mass fraction of 0.005 (yellow). The $^{18}$O is created in 
hotspots in a thin layer at the bottom of this convective region and 
mixed both outward and inward. The peak $^{18}$O mass fraction is near 
$5\by10^{-5}$, and a contour of $10^{-8}$ (red) is shown as an indication
of the maximum extent of the mixing. Again, in the first hour of 
simulation, the mixing has worked it way through approximately half 
the region expected from the 1D model. $^{18}$O also appears to be 
working its way into the non-burning portion of the helium core. 
Some of this may result from local captures of helium on $^{14}$N nuclei, 
but the irregularity of the surface argues that downward convective overshoot 
is occurring at the base of the convective region.

\begin{figure*}
\centerline{\psfig{figure=f11.eps,height=6.0in,bbllx=0pt,bblly=0pt,bburx=500pt,bbury=420pt,clip=}}
\caption{On a slice through the center of the star during run E9, the color 
shows the mass fraction of $^{18}$O.  Also shown are yellow and red 
contour lines for specific mass fractions of $^{12}$C and $^{18}$O.}
\end{figure*}

Normal stellar evolution is governed by the slow composition change 
resulting from nuclear reactions, but the structure change at the 
peak of the helium flash is a thermal-timescale adjustment.  As energy is 
produced, the convective portion of the core is transformed by 
expansion from a degenerate configuration supported by 
Fermi pressure towards a Maxwell-Boltzmann gas.  The thin region that 
separates the outer convective shell from the hydrogen burning shell 
is pushed slowly outward, and also decompressed as the gravity is 
lowered.  The effect of this on the hydrogen burning shell is most 
clearly manifested in the sharp drop in the energy production rate 
from hydrogen burning.

\begin{figure*}
\centerline{\psfig{figure=f12.eps,height=6.0in,bbllx=0pt,bblly=0pt,bburx=600pt,bbury=600pt,clip=}}
\caption{The inner and outer dark blue circles show the extent of convection in
the 1D model. The intermediate near-circles in pale blue (with triangles), 
green (with crosses), orange (withy circles) and red (with asterisks) show
the approximate outer limits of convection in the 3D model (run E9), at times 644,
1867, 3089 and $3794\thn$s. This radius expands roughly like $t^{1/2}$, and 
can be expected to catch up with the 1D boundary in $\ts 5\thn$hr.}
\end{figure*} 

Fig 12 is another way of illustrating the growth of convection in the 
helium-burning shell (run E9). The large and small dark blue circles give 
the limits of convective mixing in the original 1D model. Pale blue, green, 
orange and red circles indicate roughly the limits of convective mixing 
(according to the $^{18}$O contour) at four epochs ranging from 644$\thn$s 
to 3974$\thn$s. The inner boundary is almost independent of time, but shows 
a very slight tendency to `downwards overshoot'.
The outer boundary moves outwards, and can be approximately represented by
$r\propto t^{1/2}$. We can expect that it may reach roughly the neighbourhood
of the 1D boundary in about 5 hours.

As the energy production rates agree reasonably well between 
the 1D and 3D simulations, it is no surprise that the core expansion 
rates are similar. Over the course of an hour of simulation, the helium 
core (defined from the hydrogen composition profile) of run E9 
expanded at a rate of 18 m/s in the 3D simulation.  Beginning with 
the same model, the 1D helium core of the 1D models expanded at a 
rate of 13 m/s over a 4-hour period. While the expansion 
rate is slightly faster in the 3D simulation, the speed is far below 
the local sound speed, and hydrostatic modeling should capture most 
of the behavior of the helium flash.

Before turning to a discussion of these results, we mention a final 
numerical experiment related to the stability of helium flash 
simulations.  This experiment was preformed on the lowest luminosity 
model (E4) with the lowest mesh resolution.  A short simulation was 
done with E4 in which the mesh was not so concentrated in the helium 
burning convective shell.  Further, the ALE option was set to allow 
the mesh to slowly move into the outer regions.  The result of these 
choices is shown in Fig. 13.

\begin{figure*}
\centerline{\psfig{figure=f13.eps,height=4.0in,bbllx=0pt,bblly=0pt,bburx=600pt,bbury=420pt,clip=}}
\caption{Inadequate resolution results in greater fluctuation in 
the energy production rate, and a nuclear run-away.}
\end{figure*}

As discussed in the comparison of the energy production rate for 
models E9 and E9$\pr$, higher resolution appears to reduce the rate 
fluctuations associated with hot spots at the base of the convection zone. 
Sufficient reduction in resolution results in a nuclear run-away in 
which 100 years of energy production rate increase occurs in 10 
minutes.  Various tests done here demonstrate that all of our models 
were sufficiently resolved to eliminate this gross instability. We
believe this confirms that 3D modeling, given sufficient resolution,
is able to give convective motion that is adequate to carry the heat
flux, and does not need to be supported by (or opposed by) approximate
modeling of small-scale turbulence.

 Fig 14 shows the radial velocity color-coded so that red is outward 
and blue is inward.  The dark blue circle is the hydrogen-burning shell. 
It can be seen that there is 
possibly significant motion of an apparently convective nature {\it outside} 
this shell. It is not clear what this is due to, but it is difficult
to see why this should be only a response to helium ignition.
If it is real, it may indicate a kind of motion that might take place
above the hydrogen shell even if there is {\it no} helium flash.
Astrophysicists have frequently noted (e.g. Ivans \etal 2001, Cavallo 
\& Nagar 2000) that some giants, not necessarily
beyond the helium flash, have anomalous surface abundances that suggest
the possibility of mixing between the convective envelope and the
hydrogen-burning shell. We appear to be seeing some motion which might,
if it occurs in a pre-flash red giant, persist and grow so as to allow 
some mixing of hydrogen-burning products out to the convective envelope.
\begin{figure*}
\centerline{\psfig{figure=f14.eps,height=7.0in,bbllx=0pt,bblly=-10pt,bburx=600pt,bbury=750pt,clip=}}
\caption{Convective and other motion in run E9, about 4000s after
the start. The color-coded variable is the radial velocity, with
red outward and blue inward. The hydrogen-burning shell is marked by a
heavy narrow blue ring. Some motion is visible {\it outside}
the H-shell, in addition to the He driven convection well inside it.}
\end{figure*}

\section{Discussion}

1) The 3D simulations were robustly stable, and, apart from the 
convective shell itself, the behavior of the star was consistent with 
hydrostatic modeling, even at the peak of the helium flash. Although
exact spherical symmetry is obviously required in the 1D code, the
3D models seem to retain more-or-less spherical symmetry: there is
no tendency for one hot spot to erupt and then dominate the shell,
rendering it very asymmetric.

2) Convection is a critical element in determining the evolution 
through the helium flash.  In 1D hydrostatic modeling, convection is an 
approximation with effectively no information on the complex process 
by which hot spots develop and relax themselves. 

3) In all of our models, the convection approached but never 
exceeded the outer boundary of convection as determined from a 
stability criterion in the 1D code.  However, we cannot claim that 
overshoot will not occur in longer runs.  In 1D, the inner boundary 
of the convective shell is nearly coincident with the peak energy 
producing shell. Our simulations do show a slight and potentially
significant mixing below the 
convective shell from downward overshoot.  This leads to erosion of the 
non-burning central core, and if it continues could reduce or 
eliminate the mini-flashes that occurred in the 1D simulation. 
We intend to explore this further.

4) In the future, we intend to address all of the following: rotation,
magnetic fields, and low metallicity. We shall also pursue further the 
possibility that some slow mixing outside the hydrogen-burning shell
during First-Giant-Branch evolution might affect the surface abundances.

\section{Acknowledgments}
This study has been carried out under the auspices of the U.S. 
Department of  Energy, National Nuclear Security Administration, by 
the University of  California, Lawrence Livermore National Laboratory, 
under contract  No. W-7405-Eng-48. JL was partially supported by the 
Australian Research Council. We are indebted to R. Palasek for assistance 
with the graphics.

\end{document}